\documentclass[authoryear,preprint,12pt]{elsarticle}
\usepackage{amssymb}
\usepackage{multirow}
\usepackage{longtable}
\usepackage{url}
\usepackage{footnote}
\journal{New Astronomy}

\begin{document}

\begin{frontmatter}

\title{Optical polarization observations\\ with the MASTER robotic net}

\author[els,rvt]{M.~V.̃~ Pruzhinskaya\corref{cor1}}
\ead{pruzhinskaya@gmail.com}
\author[focal]{V.~V.̃~ Krushinsky}
\ead{krussh@gmail.com}
\author[els]{G.~V.̃~ Lipunova}
\ead{galja@sai.msu.ru}
\author[els,e]{E.~S.̃~ Gorbovskoy}
\ead{gorbovskoy@gmail.com}
\author[els]{P.~V.̃~ Balanutsa}
\ead{bala55@mail.ru}
\author[els]{A.~S.̃~ Kuznetsov}
\ead{ingell@bk.ru}
\author[els]{D.~V.̃~ Denisenko}
\ead{d.v.denisenko@gmail.com}
\author[els]{V.~G.̃~ Kornilov}
\ead{victor.kornilov@gmail.com}
\author[els]{N.~V.̃~ Tyurina}
\ead{tiurina@sai.msu.ru}
\author[els,rvt,e]{V.~M.̃~ Lipunov}
\ead{lipunov2007@gmail.com}
\author[rv]{A.~G.̃~ Tlatov}
\ead{tlatov@mail.ru}
\author[rv]{A.~V.̃~ Parkhomenko}
\ead{gas-ran@mail.ru}
\author[r]{N.~M.̃~ Budnev}
\ead{nbudnev@api.isu.ru}
\author[r]{S.~A.̃~~ Yazev}
\ead{uustar@star.isu.ru}
\author[r]{K.~I.̃~ Ivanov}
\ead{ivorypalace@gmail.com}
\author[r]{O.~A.̃~ Gress}
\ead{grol08@rambler.ru}
\author[rt]{V.~V.̃~ Yurkov}
\ead{stella@ascnet.ru}
\author[rt]{A.~V.̃~ Gabovich}
\ead{stella@ascnet.ru}
\author[rt]{Yu.~P.̃~ Sergienko}
\ead{sergienkoyp@mail.ru}
\author[rt]{E.~V.̃~ Sinyakov}
\ead{marginalis44@yandex.ru}

\cortext[cor1]{Corresponding author}

\address[els]{Lomonosov Moscow State University, Sternberg Astronomical Institute, Universitetsky~pr.,~13,~Moscow, Russia, 119991}
\address[rvt]{Faculty of Physics, Lomonosov Moscow State University, Moscow, Russia}
\address[e]{``Extreme Universe Laboratory'' of Lomonosov Moscow State University Skobeltsyn Institute of Nuclear Physics, Moscow}
\address[focal]{President Yeltsin Ural Federal University, Ekaterinburg, Russia}
\address[rv]{Mountain Astronomical Station, Main (Pulkovo) Astronomical Observatory, Russian Academy of Sciences, St. Petersburg, Russia}
\address[r]{Irkutsk State University, Irkutsk, Russia}
\address[rt]{Blagoveshchensk State Pedagogical University, Blagoveshchensk, Russia}

\begin{abstract}
We present results of optical polarization observations performed with the MASTER robotic net \citep{master, master1, ExpA} for three types of objects: gamma-ray bursts, supernovae, and 
blazars.
For the Swift gamma-ray bursts GRB100906A, GRB110422A, GRB121011A,  polarization observations were obtained during very early stages of optical emission. For GRB100906A it was the first prompt 
optical polarization 
observation in the world. 
Photometry in polarizers is presented for Type Ia Supernova 2012bh during 20 days, starting on March 27, 2012. We find that the linear polarization of SN 2012bh at the 
early stage of the envelope expansion was less than 3\%. 
Polarization measurements for the blazars OC 457, 3C 454.3, QSO B1215+303, 87GB 165943.2+395846 at single nights are presented.
We infer the degree of the linear polarization and polarization angle.
The blazars OC 457 and 3C 454.3 were observed during their periods of activity. 
The results show that MASTER is able to measure substantially polarized light; at the same time it is not suitable for determining weak polarization (less than 5\%) of dim objects
(fainter than $16^m$). 
Polarimetric observations of the  optical emission from gamma-ray bursts and supernovae are necessary to investigate the nature of these transient objects.
\end{abstract}


\end{frontmatter}

\section{Introduction}
Polarimetry plays an important role in modern astrophysics.  
Polarization measurements provide information about the nature of radiation sources, geometrical properties of the emitting regions, the spatial distribution of matter around sources, magnetic fields.

In the last two decades, the polarimetry, especially spectropolarimetry, has advanced a lot. Different polarization techniques and devices have been developed. Fast CCD cameras and new polarizing 
materials made  polarimetric observations possible for a rising number of optical telescopes.
In particular, polarization measurements are important for short-living or fast variable objects such as gamma-ray bursts (GRBs) and supernovae. 

The first polarization measurements of GRBs afterglow  were realized by large telescopes: the 8.2-m Unit Telescopes of VLT 
\citep{Covino 1999, Rol 2000, Covino 2002}, the 10-m Keck I telescope 
 \citep{Barth 2003}, the 6.5-m Multiple Mirror Telescope \citep{Bersier 2003}. It's remarkable that the first detections of the early optical polarization were performed by a robotic 
telescope: the 2-m robotic Liverpool telescope
 \citep{Mundell 2007, Steele 2009}.
There are also several telescopes with average-size aperture that measure GRBs polarization: the 2.56-m Nordic Optical Telescope 
 \citep{Rol 2003, Greiner 2003}, the 2.2-m telescope of Calar Alto Observatory 
 \citep{Greiner 2003}, 
the Kanata 1.5-m telescope at Higashi-Hiroshima Observatory 
 \citep{Uehara 2012}, and the 1-m USNO telescope in Flagstaff \citep{Greiner 2003}.

The early attempts to measure polarization of Type Ia supernovae were made by using the 2.6-m Shain reflector built for the Crimean Astrophysical Observatory \citep{Shakhovskoi 1976}. 
The first spectropolarimetric data were obtained 
on the 3.9-m Anglo-Australian Telescope \citep{McCall 1984}. 
The program of polarimetric observations of SN Ia was begun on the 2.1-m Struve Telescope at McDonald Observatory 
\citep{Wang 1996, howell}. This program was continued on the 8.2-m unit telescopes of the VLT  \citep{Wang 2007}.
Spectropolarimetry is also carried out on the Keck I 10-m telescope  \citep{chornock}.

The first MASTER telescope was mounted in 2002 near Moscow. 
The MASTER net began operating in full mode in 2010 \citep{master, master1, ExpA, Gorb}. 
More than 100 alert pointings at GRBs were made by MASTER. The MASTER net holds
the first place in the world in terms of the total number of first pointings, and currently more than a half of
first pointings at GRBs by ground telescopes are made by the MASTER net. 
More than 400 optical transients have been discovered, among them cataclysmic variables, supernovae, blazars, potentially hazardous asteroids, transients of unknown nature. 
Photometry of 387 supernovae has been carried out.
Polarization measurements is one of the purposes for us to design and construct the MASTER II robotic telescopes. 
The  linear polarization
measurements with the MASTER net involve simultaneous observations of an object by several telescopes equipped with cross linear polarizers. Since each telescope of the net has two 
wide-FOV astronomical tubes, we have to  point at least two telescopes\footnote{The photometer in Kislovodsk contains 4 differently oriented polarizers, which 
allows us to measure linear polarization at this site alone.} to an object simultaneously in order to determine its Stokes parameters.

The unique design of MASTER II makes it the only wide-FOV fully robotic instrument in the world able to measure polarization. In this work we report the results of the examination of its accuracy   and 
analyze its capability to measure polarization of different types of astrophysical objects: gamma-ray bursts, supernova, blazars.

\section{MASTER instruments and reduction of observations}
Each MASTER II telescope contains a two-tube aperture system with a total field of view of 8 square degrees, equipped with a 4000~pixel~$\times$~4000~pixel CCD camera with a scale
of $1.85''$/pixel, an identical photometer with B, V, R, and I filters representing the Johnson-Cousins system, and polarizing filters. Both optical tubes are installed on a high-speed
mount with position feedback, which does not require additional guide instruments for exposures not exceeding three to ten minutes. The setup has an additional degree of
freedom: the variable angle between the optical axes of the two tubes. 
This allows us to double the field of view during survey observations if the tubes are deflected from each other, or to conduct synchronous multi-color  photometry with parallel tubes. 
A single telescope of the MASTER II net provides a survey speed of 128 square degrees per hour with a limiting magnitude of $20^m$ on dark moonless nights.

Astrometric and photometric calibration is made by a common method for all MASTER observatories \citep{Gorbovskoy 2012, ExpA}.
Bias and dark subtraction, flat field correction, and astrometry
processing are made automatically. Bias and dark images obtained before the beginning of observations and the closest in time flat field images, obtained on the twilight sky, are used. 
To convert magnitudes into absolute fluxes, zero points of the MASTER bands should be used. They can be found at \url{http://master.sai.msu.ru/calibration/}.

The first MASTER polarizers were high contrast Linear Polarizing Films combined with usual glass (in January--July 2011  they were combined with R filter instead).
Since July 2011 all polarizers have been replaced by new broadband polarizers manufactured using linear conducting nanostructure technology  \citep{ExpA, Seh}. 
Magnitudes obtained from broadband photometry correspond to $0.2B + 0.8R$ where $B, R$ are the standard Johnson filters \citep{Gorbovskoy 2012}.
Each tube is equipped with one polarizer, the polarization directions of two tubes in a single assembly being perpendicular to one another.
Polarizers' axes are set in two ways with respect to celestial sphere: in the MASTER Kislovodsk and the MASTER Tunka sites, the axes are directed at positional angle $0^\circ$ and $90^\circ$
 to the celestial equator (polarizers oriented at $45^\circ$ and $135^\circ$ were added in Kislovodsk in April 2012),
in the MASTER Blagoveshchensk and the MASTER Ural, at angles $45^\circ$ and $135^\circ$. Thereby, using several MASTER telescopes one can 
realize observations with different orientation of the polarizers. Such a construction is very effective for fast events with significant intrinsic polarization. These events are mostly of 
extragalactic origin. In spite of the fact that all MASTER telescopes have a similar construction, optical scheme, and polarizers, there are some uncertainties in the channels responses. 
These limit the precision of polarization measurements.
Also, the calibration using known polarized Galactic sources is not possible, as any measurement of polarization involves  subtraction of nearby field stars between at 
least three images, obtained with tubes with differently oriented polarizers. Thus the derived polarization of a target object is always relative to the nearby field stars. 
They are usually chosen from the 
area around the target source  in a radius of $10'$ and bound to have the same polarization due to the similar interstellar matter properties toward them.  
The original stellar radiation is not polarized, polarization appears when light 
passes through the interstellar dust. Interstellar polarization can reach high values and strongly depends on the Galactic direction and wavelength \citep{ISP}.

The goals of the MASTER polarimetric analysis, 
when observing highly-polarized extragalactic events, are: to 1) find polarization of an event in excess of the average polarization of  Galactic stars in that direction, 
2)  remove additional systematic polarization introduced by the instruments, 3) estimate the uncertainty of polarimetric measurements by MASTER.

Stars with zero polarization are required for the channel calibration. We assume the polarization of light from stars  in the field of view is small. This can be checked using Serkowski law \citep{Serkowski1975}. 
The difference in magnitudes between two polarizers orientations 
averaged for all reference stars gives the correction that takes into account different channel responses. 

\section{GRB polarization observations}
Gamma-ray bursts are the most powerful explosions in the Universe. Unfortunately, the physics of the process is not fully understood. In particular, this is due to the fact that GRBs are 
short-living events. The spectral investigations argue for synchrotron nature of the GRB emission. It is known that the synchrotron emission in ordered magnetic field is polarized. 
Thus, observable polarization depends on the degree of coherence of magnetic fields and on the geometrical properties of the emitting regions.
Theoretical models with ordered magnetic field predict polarization about 20-30\% \citep{Granot}. It is a challenge to test this prediction since  GRB are fast variable objects 
 with characteristic times about tens of seconds. 

We formally divide optical observations of GRBs into three types: afterglows, early-time, and prompt, in succession of time elapsed from the gamma trigger. Historically, the optical 
observations of GRBs progressed from afterglows to prompt.

To date, there are no robust positive observations of GRB prompt emission polarization, and only several measurements of afterglow polarization, 
generally on a level of about 1-3\%. These are GRB990510 that showed $(1.7 \pm 0.2)\%$ polarization at 
around $980~T_{90}$ \citep{Covino 1999}, GRB020813, with detected polarization varying from 2\% ($670-1140~T_{90}$)  to 0.8\%  ($3300-3570~T_{90}$) \citep{Barth 2003, Covino 2002},
GRB021004  with $\sim 1-2$\% at  $t>500~T_{90}$ \citep{Lazzati 2003} and GRB030329 with $0.3-2.5$\% ($t>1700~T_{90}$) \citep{Greiner 2003}\footnote{Here, $T_{90}$ is the standard 
characteristic time for GRB duration, whose values can be found in the papers cited above.}.
An exception is GRB020405 for which a highly significant polarization $(9.9 \pm 1.3)\%$ (1.3 days after the GRB) was detected \citep{Bersier 2003}. However this result is not confirmed by 
measurements of other teams, which obtained $(1.5 \pm 0.4)\%$ (1.2 days after the GRB) \citep{Masetti 2003}; $(1.96 \pm 0.33)\%$ and $(1.47 \pm 0.43)\%$ 
(2.2 and 3.25 days after the GRB) \citep{Covino 2003}.

For afterglow times,  we can probe the physics of a relativistic forward shock expanding into surrounding media. 
Characteristic times can help to distinguish the nature of polarization variability as intrinsic or interstellar (see \cite{Covino 2004}). 
Dust destruction in the vicinity of a gamma-ray burst due to the strong radiation field \citep{Waxman Draine 2000, Perna Lazzati 2002} would result in a monotonically decreasing 
polarization degree with a constant position angle \citep{Greiner 2003}. For such low levels of polarization, the analysis is difficult, and distinguishing the intrinsic polarization from 
the polarization in  the ISM  of Milky Way and host galaxy is model-dependent. 

Thus, much effort was put to manage early-time observations, when polarization of the GRB-emission itself  is anticipated to be more pronounced.  The origin of early optical radiation is also
 considered alternatively: it can be from the starting forward shock or from the reverse shock \citep{Sari 1998, Sari Piran 1999, Kobayashi 2000}.
In the latter case, polarization is expected to manifest the structure of intrinsic magnetic filed  more directly. 

At the front of the forward shock, patches of coherent magnetic filed may be generated stochastically.  A theoretical model by \cite{Gruzinov Waxman 1999} suggests  10\% polarization as 
an upper level with erratic variations of the position angle. Specific patterns of polarization evolution are predicted in the models concerning beamed synchrotron  radiation from a jet 
observed out of its symmetry axis, even if magnetic fields are completely tangled (see references in the review by \cite{Covino 2004}). So far, unfortunately, predicted patterns are not 
seen clearly in  observations, implicating that the presumed jet models are far too simple.

Few early observations of optical polarization exist.  For GRB060218, \citet{Mundell 2007} reported an upper limit of 8\% for times corresponding to the onset of the forward shock, 
basing on a 30 s-exposure observation started at $2~T_{90}$. GRB090102 \citep{Steele 2009} manifested 10\% polarization in the early optical emission (a single 60 s exposure started at
 $6~T_{90}$), with emission interpreted as an emission from a reverse shock.  Both polarization measurements were performed with a rotating polarizer of the 2.0-m Liverpool robotic 
telescope.  GRB091208B \citep{Uehara 2012} revealed ~10\% in optical emission, interpreted as an early afterglow (forward-shock emission) obtained on the Hagashi-Hiroshima 1.5m telescope,
 measured for an interval $10-50~T_{90}$.

The prompt observations, simultaneous with the gamma signal, may shed light on another physics, more directly connected with the GRB engine -- the structure of the  magnetic field within the 
original outflow. Prompt GRB emission is thought to originate from the jet material interacting with itself.  The magnetic fields can be advected from a central engine. If the fireball is
 electromagnetically dominated, during the first ten minutes the degree of polarization may be $>$40\% \citep{Lazzati 2006}. 

For the prompt GRB emission, so far only observations in gamma diapason delivered polarization results: first, GRB021206 was claimed to have 80\% polarized prompt gamma-emission but this was
 critisized later (see \cite{Covino 2004}).  Later, several GRBs are reported with linear polarization of gamma prompt emission at levels 20-80\%:  GRB041219 \citep{Gotz 2009},
 GRB100826A \citep{Yonetoku 2011}, and, recently,  GRB110301A  and GRB110721A  \citep{Yonetoku 2012}. 

We would like to stress that especially in cases when optical flash is observed simultaneously with the gamma-ray emission, detection of the polarization of the GRBs optical 
counterparts is a crucial component to our understanding of the jet physics. If it is produced by a reverse shock, intrinsic magnetic fields can be tested. 

The MASTER net was designed with the objective to deliver polarization information as early as possible after GRB triggers.
More than 100 observations of GRB were made by the MASTER global robotic net. Optical emission was detected for 11 GRBs. 
%On each telescope observations are carried out by two tubes with mutually orthogonal polarizers.
GRB100906A, GRB110422A and GRB121011A deserved attention because their optical observations were carried out during the gamma-ray emission.
Regarding the three GRBs considered below, the optical polarization observations  were made by MASTER only in two polarizers, and no significant difference was found in two channels for each of them. 
Unfortunately, it is impossible to make a conclusion about the absence of polarization if you have 
observations only in two polarizers (there is a possibility that the polarization plane lies at 45 degrees to the polarizers). 
Electromagnetic radiation can be described in terms of the Stokes vector, which consists of 4 components: I (total intensity); Q, U (describe linear polarization);
V (describes circular polarization). The degree of linear polarization $P$ and polarization angle $\theta$ can be expressed in terms of the Stokes parameters as $P = \frac{\sqrt{Q^2 + U^2}}{I}, 
\theta = \frac{1}{2}arctan\frac{U}{Q}$.
Single dimensionless Stokes parameter is the lower limit of the degree of linear polarization, that is, the actual $P$ can be positive 
even if one Stokes parameter ($\frac {I_1 - I_2} {I_1 + I_2}$) is zero with some uncertainty. The problem of probabilistic estimates of  the degree of 
linear polarization from observations with two polarizers was addressed in the Appendix of \citet{Gorbovskoy 2012}.

\subsection{GRB100906A}

\begin{figure}[h!]\centering
\includegraphics[width=0.9\linewidth]{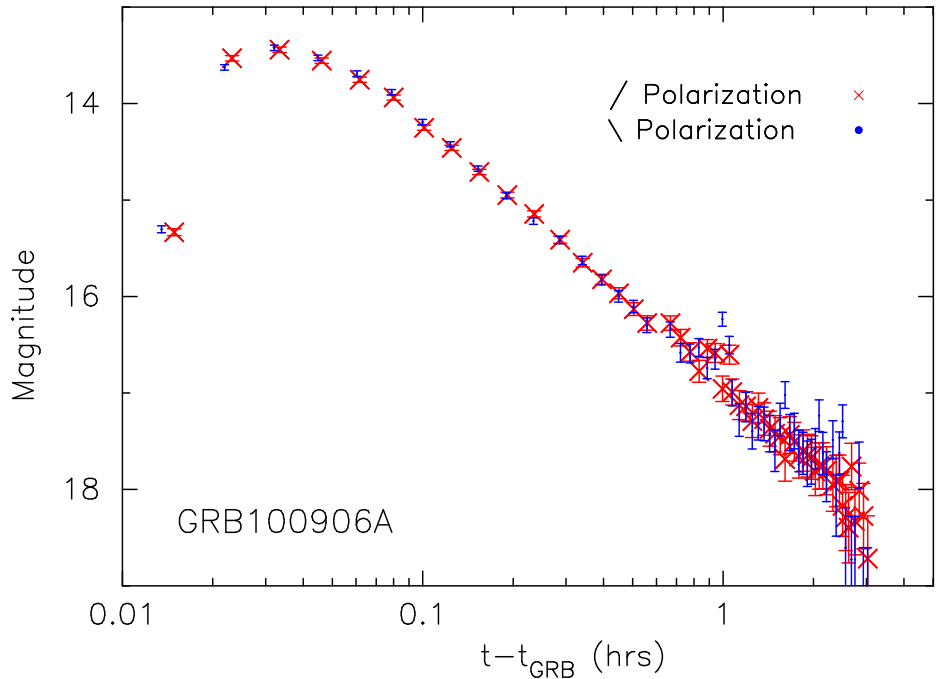}
\caption{Light curve of GRB100906A in mutually perpendicular polarizers obtained with MASTER Tunka \citep{Gorb}.}
\label{fig:lc100906} 
\end{figure}

The Tunka MASTER II telescope pointed at the object 23 s after receiving the alert signal (38 s after the registration of the burst by the Swift space observatory \citep{Evans 2010}). 
The object of $13^m$ was detected. A series of frames using polarizers with exposures growing from 10~s to 180~s were obtained over the next four hours (see table 4 of \cite{Gorbovskoy 2012}). 
The duration of the
GRB in the gamma-rays was $T_{90} = (114 \pm 1.6)$~s \citep{Markwardt 2010}. Thus, the first three frames of the series were obtained simultaneously with the GRB. For the first three minutes, 
the GRB was synchronously observed in optical, X-ray ($0.3-10$~keV), and gamma diapason ($15-150$~keV). Note that it was the first optical polarization observations of prompt GRB emission. 
The absolute photometry in MASTER wide band is presented in Fig. \ref{fig:lc100906}.
The differences in signals obtained in each polarization in the time interval from 100 s to $10^4$ s were computed ($\frac {I_1 - I_2} {I_1 + I_2}$). It was found that this difference is 
smaller than the standard deviation of $\frac {I_1 - I_2} {I_1 + I_2}$ for field stars, which is equal to 2\%. 

\subsection{GRB110422A}
\begin{figure}[h!]\centering
\includegraphics{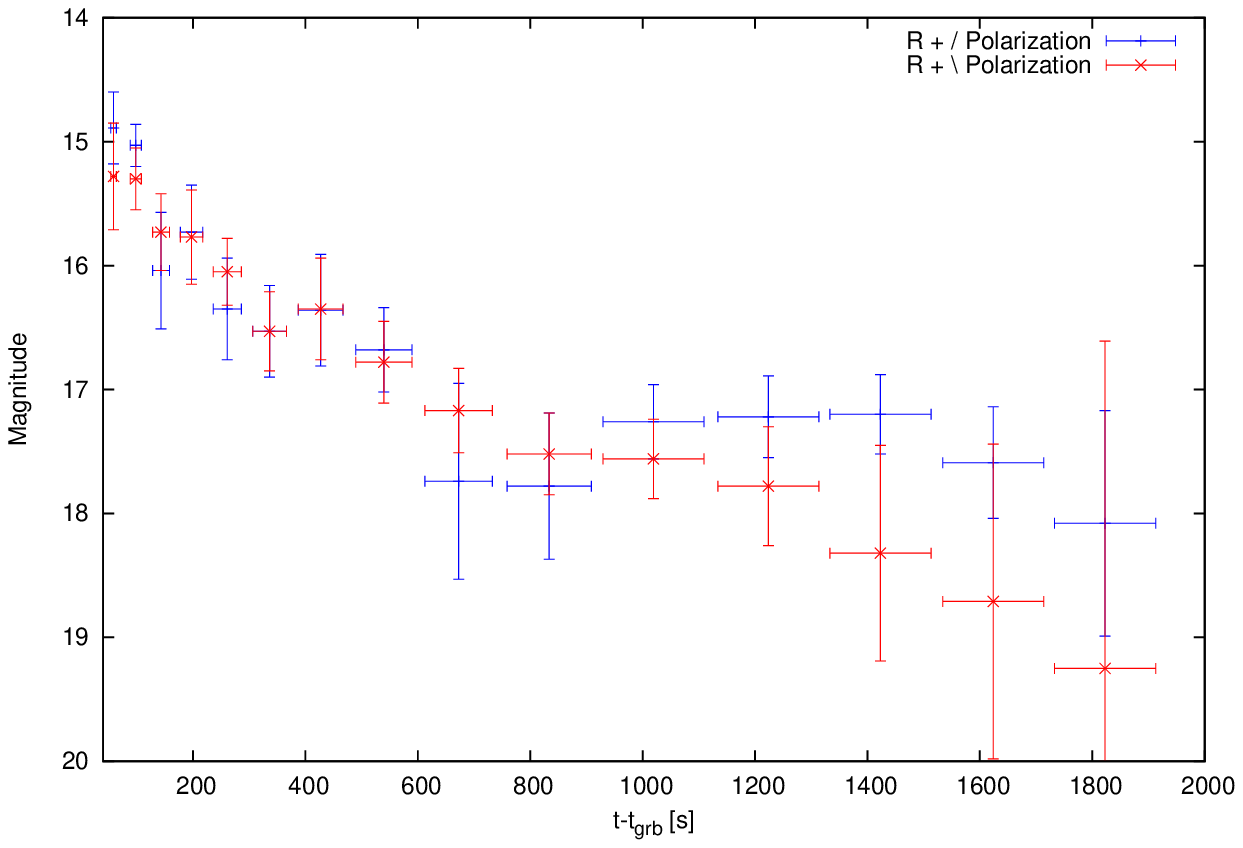}
\caption{Light curve of GRB110422A in mutually perpendicular polarizers and R filter obtained with MASTER Tunka \citep{Gorb}.}
\label{fig:lc110422} 
\end{figure}

The Tunka MASTER telescope was the first ground-based telescope to point at GRB110422A \citep{Mangano 2011}, 53~s after beginning of the burst and 37~s after receiving the alert. 
The observations were carried out in both tubes in mutually perpendicular polarizations. The exposures were successively increased from 10~s to 180~s. An optical transient was detected 
in the very first 10-s frame. The observations were disrupted 35 min after they started because of bad weather conditions. 30 exposures were obtained, 15 in each polarization 
(see Fig. \ref{fig:lc110422} and Table \ref{tab:grb110422}) \citep{Gress 2011, Gress 2}.

\begin{table}\centering
\caption{Optical observations of GRB110422A in mutually perpendicular polarizers and R filter in MASTER Tunka \citep{Gorb}. The absolute fluxes can be obtained using zero-points from http://master.sai.msu.ru/calibration/.}
\begin{tabular}{|c|c|c|c|c|}
\hline 
\label{tab:grb110422} 
$T-T0$ (s)& R+P\textbackslash{} (mag)& Err(R+P\textbackslash{}) (mag)& R+P/ (mag)& Err(R+P/) (mag)\tabularnewline
\hline 
58.5 & 15.28 & 0.43 & 14.89 & 0.29\tabularnewline
98.1 & 15.3 & 0.25 & 15.03 & 0.17\tabularnewline
143.0 & 15.73 & 0.31 & 16.04 & 0.47\tabularnewline
197.3 & 15.77 & 0.38 & 15.73 & 0.38\tabularnewline
261.0 & 16.05 & 0.27 & 16.35 & 0.41\tabularnewline
336.4 & 16.53 & 0.32 & 16.53 & 0.37\tabularnewline
427.1 & 16.35 & 0.41 & 16.36 & 0.45\tabularnewline
539.4 & 16.78 & 0.33 & 16.68 & 0.34\tabularnewline
672.5 & 17.17 & 0.34 & 17.74 & 0.79\tabularnewline
833.7 & 17.52 & 0.33 & 17.78 & 0.59\tabularnewline
1019.3 & 17.56 & 0.32 & 17.26 & 0.3\tabularnewline
1223.8 & 17.78 & 0.48 & 17.22 & 0.33\tabularnewline
1423.0 & 18.32 & 0.87 & 17.2 & 0.32\tabularnewline
1623.7 & 18.71 & 1.27 & 17.59 & 0.45\tabularnewline
1822.8 & 19.25 & 2.64 & 18.08 & 0.91\tabularnewline
\hline 
\end{tabular}
\end{table}

In spite of the fact that the GRB formally ended before the first MASTER image (the GRB duration was $T_{90}^{BAT} = (25.9 \pm 0.6)$ s (Swift/BAT, \cite{Palmer}), $T_{90}^{KW} \sim 40$ s 
(Konus/WIND, \cite{Golenetskii}), 
the burst exceeded the $3\sigma$ level 
up to 115~s in the Swift Burst Analyser data \citep{Evans 2009}. At least two MASTER observations were made during this interval. Thus MASTER detected prompt emission at those times.
The GRB redshift, z = 1.77, was independently determined by different groups \citep{Malesani,Postigo}.
The dimensionless Stokes parameter does not exceed that of field stars and is less than 2\%.

\subsection{GRB121011A}
\begin{figure}[h!]\centering
\includegraphics[width=0.9\linewidth]{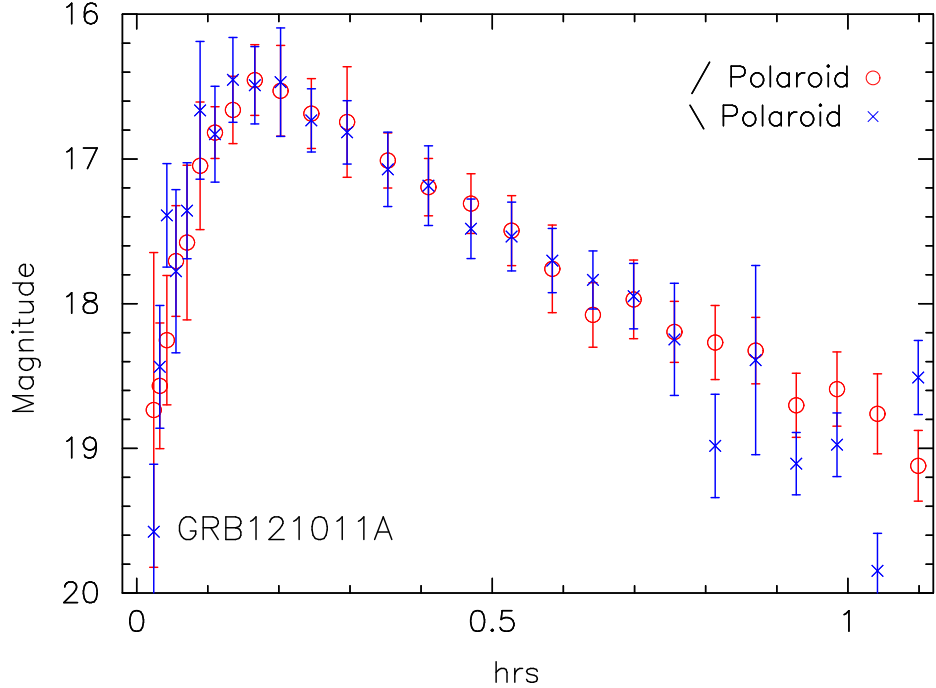}
\caption{Light curve of GRB121011A in mutually perpendicular polarizers obtained with MASTER Blagoveshchensk.}
\label{fig:lc121011a} 
\end{figure}

GRB121011A  \citep{gcn13845, gcn13859, gcn13860} was observed by the MASTER telescopes at two sites \citep{gcn13848,gcn13854,gcn13861}. The robotic telescope located in
Blagoveshchensk was pointed to GRB121011A  51 s after trigger time $T0$ in two polarizations. The Tunka MASTER~II robotic telescope (near Baykal lake) was pointed to GRB121011A 106 sec after $T0$,
 also in two polarizations.  Thus, the observations began in four polarizations. Unfortunately, the weather conditions  turned out  poor at Tunka site. The first individual image, on
 which the optical counterpart was found with Tunka telescope, was taken  230 s after $T0$.  About $35$~min of optical observations were obtained with the Blagoveshchensk telescope of the
 MASTER net. The light curve is shown in Fig.~\ref{fig:lc121011a} and photometry is presented in Table \ref{tab:grb121011A}. 

\begin{longtable}[H]{|c|c|c|c|c|c|}
%\centering
\caption{Optical observations of GRB121011A in mutually perpendicular polarizers in MASTER Blagoveshchensk. The absolute fluxes can be obtained using zero-points from http://master.sai.msu.ru/calibration/.}\\
%\begin{tabular}{c|c|c|c|c|c} 
\hline 
\label{tab:grb121011A} 
$T-T0$ (hrs)&Exposure (s)&P\textbackslash{} (mag)& Err(P\textbackslash{}) (mag)& P/ (mag)& Err(P/) (mag)\tabularnewline
\hline 
0.0240  &  20  &  19.6  &  0.5  &  18.7 &   1.1  \\  
0.0326  &  20  &  18.4  &  0.4  &  18.6 &   0.4  \\
0.0426  &  30  &  17.4  &  0.4  &  18.3 &   0.4  \\  
0.0553  &  40  &  17.8  &  0.6  &  17.7 &   0.4  \\  
0.0708  &  50  &  17.4  &  0.3  &  17.6 &   0.5  \\  
0.0892  &  60  &  16.7  &  0.5  &  17.0 &   0.4  \\
0.1102  &  70  &  16.8  &  0.3  &  16.8 &   0.2  \\
0.1354  &  90  &  16.5  &  0.3  &  16.7 &   0.2   \\ 
0.1662  &  110 &   16.5 &   0.3 &   16.5&    0.2   \\ 
0.2024  &  130 &   16.5 &   0.4 &   16.5&    0.3    \\
0.2457  &  160 &   16.7 &   0.2 &   16.7&    0.2   \\ 
0.2959  &  180 &   16.8 &   0.2 &   16.7&    0.4   \\ 
0.3531  &  180 &   17.1 &   0.3 &   17.0&    0.2   \\ 
0.4103  &  180 &   17.2 &   0.3 &   17.2&    0.2   \\ 
0.4701  &  180 &   17.5 &   0.2 &   17.3&    0.2   \\ 
0.5272  &  180 &   17.5 &   0.2 &   17.5&    0.2    \\
0.5843  &  180 &   17.7 &   0.2 &   17.8&    0.3   \\ 
0.6416  &  180 &   17.8 &   0.2 &   18.1&    0.2    \\
0.6987  &  180 &   17.9 &   0.2 &   18.0&    0.3    \\
0.7560  &  180 &   18.2 &   0.4 &   18.2&    0.2    \\
0.8133  &  180 &   19.0 &   0.4 &   18.3&    0.3    \\
0.8703  &  180 &   18.4 &   0.7 &   18.3&    0.2    \\
0.9273  &  180 &   19.1 &   0.2 &   18.7&    0.2    \\
0.9845  &  180 &   19.0 &   0.2 &   18.6&    0.3    \\
1.0416  &  180 &   19.8 &   0.3 &   18.8&    0.3    \\
1.0989  &  180 &   18.5 &   0.3 &   19.1&    0.2    \\
1.1562  &  180 &   19.0 &   0.3 &   18.4&    0.4    \\
1.2134  &  180 &   18.2 &   0.3 &   19.0&    0.1    \\
1.2706  &  180 &   19.5 &   0.3 &   18.8&    0.3    \\
1.3277  &  180 &   19.7 &   0.2 &   19.2&    0.2    \\
1.3848  &  180 &   18.7 &   0.3 &   18.5&    0.2    \\
1.4421  &  180 &   21.4 &   0.5 &   19.5&    0.3    \\
1.4996  &  180 &   19.2 &   0.3 &   19.4&    0.3    \\
1.5566  &  180 &   19.3 &   0.3 &   19.1&    0.2    \\
1.6139  &  180 &   19.5 &   0.7 &   20.2&    0.3    \\
1.6711  &  180 &   18.9 &   0.3 &   19.1&    0.3    \\
1.7854  &  180 &   21.2 &   0.3 &   19.0&    0.2    \\
1.9324  &  180 &   18.9 &   0.3 &   21.6&    0.4    \\
2.0145  &  180 &   19.8 &   0.4 &   20.5&    0.3    \\
2.0717  &  180 &   21.1 &   0.3 &   19.9&    0.3    \\
2.1289  &  180 &   20.9 &   0.2 &   19.6&    0.3    \\
2.1861  &  180 &   20.5 &   0.3 &   21.9&    0.3    \\
2.2434  &  180 &   19.6 &   0.5 &   23.5&    0.2    \\
2.3008  &  180 &   21.0 &   0.6 &   19.9&    0.3    \\
2.4152  &  180 &   19.7 &   0.3 &   19.7&    0.3    \\
2.4725  &  180 &   18.8 &   0.4 &   19.3&    0.2    \\
2.5296  &  180 &   19.1 &   0.4 &   19.6&    0.4    \\
2.5617  &  180 &   20.9 &   0.5 &   21.9&    0.4    \\
2.7013  &  180 &   19.5 &   0.5 &   18.8&    0.4    \\
\hline 
%\end{tabular}
% \label{tab:grb121011A} 
\end{longtable}

For interval 0.1-1 hr after $T0$, we obtain a zero time-averaged Stokes parameter: $\frac {I_1 - I_2}{I_1 + I_2}=(1\pm2)\%$.
For the derived uncertainty of 2\%, 1-$\sigma$ upper limit for the degree of linear polarization $P$ is about 15\% (see Fig.
14 of Gorbovskoy et al 2012: value of $P$=15\% matches 1-$\sigma$ probability $\mathcal{L}=100-68\% = 32\%$ for the curve corresponding to a relative accuracy 2\%).
Another characteristic value is the standard dispersion of Stokes parameters of field stars. It describes the ultimate
accurcy of determination of the Stokes parameter  for the specific line of sight in the sky at the time. We obtain 6\% as the characteristic value of this dispersion for observations 
of GRB 1210011A.

\section{Supernovae polarization observations}

As precise standardizable candles, Type Ia supernovae have been used to trace the expansion of the Universe as a function of redshift, leading to the discovery of the accelerated expansion 
\citep{Perlmutter, Riess}. 
However, several questions related to the physics of the explosion mechanisms are still not well understood. 
It is generally believed that there are two main mechanisms of Type Ia supernovae explosions: the merger of two white dwarfs \citep{DD, DD2} and the Schatzman mechanism, according to which 
a burst is 
a result of matter accretion on a white dwarf from the companion star in binary systems \citep{SD}. 
In the first case, which is more often realized in elliptical galaxies \citep{lipunov}, the specific angular momentum of matter is higher than in the second case. This could lead to 
an anisotropy of the explosion and asymmetrical loss of envelope and, consequently, a significant polarization of optical radiation  is expected. Thereby, 
the registration of significant (more than 2\%) polarization can be an independent argument for a model of merging white dwarfs.
It should be noted that the continuum polarization 
depends on the geometry of the explosion but line polarization associates with distribution of matter around supernovae \citep{Wang and Wheeler}.

Previous observations of Type Ia supernovae display close to zero continuum polarization and modest line polarization $\sim1\%$. Even if continuum polarization is observed, 
its value is small: for example, the continuum polarization for SN 1999by (prototype of which is SN~1991bg)   was $0.3-0.8\%$ \citep{howell}, for SN 2005hk $\sim0.4\%$ \citep{chornock}.
 
SN 1996X was the first SN Ia with spectropolarimetry prior to the optical maximum. The broadband polarimetry showed that continuum polarization is zero. The spectropolarimetry demonstrated 
spectral features with a rather low polarization $\sim0.3\%$ \citep{wang}. 

It's very important to measure polarization before optical maximum while an envelope is not expanded essentially. Polarization at the latest phase of any supernova is low. For example, 
 \cite{leonard} report spectropolarimetric observations of SN 1997dt 21~days after optical maximum. Polarization was not detected but inessential line polarization in FeII 
and SiII was found. Another example is SN 2001el \citep{wang2003a}. Before optical maximum the linear polarization in the continuum was $\sim0.2-0.3\%$. 
During the next 10~days the degree of continuum and line polarization decreased and disappeared entirely~$\sim$~19~days after the optical maximum. Spectropolarimetry of SN~2004S~9~days 
after the maximum light displayed very low polarization \citep{chornock2006}. The absence of polarization in the later stages stresses the importance of early polarimetric observations.

Significant line polarization was found for SN 2004dt, for which there were data approximately 7 days before  \citep{wang2006} and 4 days after the maximum light \citep{leonard}. 
During this period the polarization of the SiII line varied within the range $\sim2\%$. Measurements of polarization of the SN 2002bf at approximately the similar time interval as for the SN 2004dt
showed CaII line polarization $\sim2\%$ \citep{leonard}. SN 1997bp and SN 2002bo also showed $1-2\%$ line polarization \citep{wang2006}.

Interstellar polarization (ISP) in our Galaxy or in host galaxies of supernovae  complicates determination of intrinsic supernova polarization. ISP can be estimated using the
empirical Serkowski law \citep{Serkowski1973}. This law was obtained based on observations in our Galaxy, thus we cannot be certain that distribution and structure 
of dust in host galaxies of supernovae are the same. Host-galaxy ISP can be high enough. For example, in SN 2006X  
polarization uncorrected for ISP declines from $8\%$ at 4000\AA ~to  $\sim2\%$ at 8000\AA ~\citep{Patat}. Taking into account ISP these authors obtain polarization of the CaII IR line $\sim1.5\%$ and 
the SiII line $\sim0.5\%$ (10 days before the maximum light).

The number of Type Ia supernovae detected before maximum light is small but a number of those, for which spectropolarimetric or/and polarimetric data were obtained, is much less.
SN 2012bh is a good example of a Type Ia supernova that was discovered before maximum light.
SN 2012bh, exploded in the Sb galaxy UGC 7228, was first detected by the Pan-STARRS1 Medium Deep Survey on 2012 March 11.50 UT  
\citep{PanStar, CBET3066, Atel4293}. The coordinates of the supernova are $R.A.=12^h13^m37^s.309, Decl.=+46^{\circ}29'00''.48$ (J2000.0). 
The object was discovered with an approximate brightness of $z_{P1} = 22.8 (AB)$ mag and brightened to $i_{P1} = 16.99$ mag by 2012 March 21.47 UT \citep{PanStar}. 
The spectrum in the range 480-940 nm obtained on March 15  with the Grand Telescope Canaries identified the object as a young normal SN Ia. 
Another spectrum in the range 350-740 nm was obtained on March 23 with the 1.5-m telescope at the Fred~L.~Whipple Observatory on Mount Hopkins, Arizona.
Joint processing of these two spectra showed that supernova was detected three weeks prior to the maximum light.

\begin{figure}[h!]\centering
\includegraphics[width=0.6\linewidth]{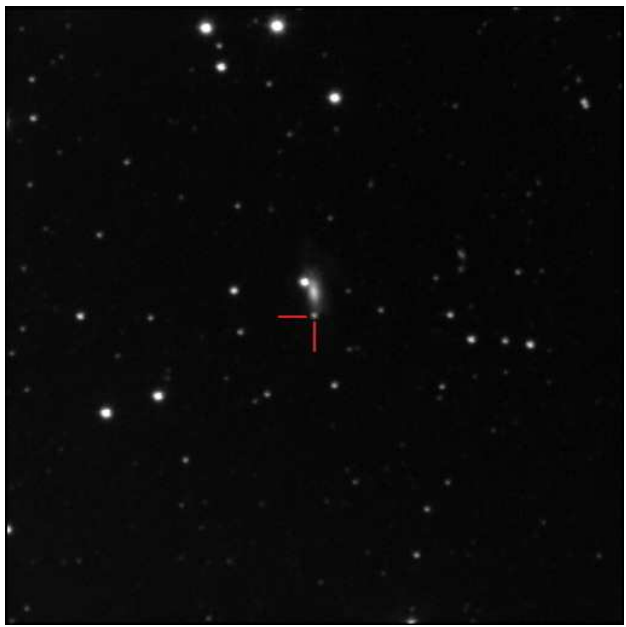}
\caption{Supernova 2012bh observed by MASTER Tunka on April 1, 2012.}
\label{ris:image} 
\end{figure}

The observations of SN 2012bh were made by the MASTER robotic net in Tunka, Kislovodsk, and Blagoveshchensk in polarizers.
MASTER observed the supernova before and after the maximum light, from March 27 to April 15.
The image of the supernova obtained by MASTER Tunka is presented in Fig.~\ref{ris:image}. 

The photometry is made in the IRAF/apphot \citep{Tody}. 
From the SDSS-DR7 catalog \citep{sdss} 43 comparison stars were selected with the help of application Aladin \citep{aladin}. All these stars are  less than $10'$ from the supernova and brighter 
than 18 mag.
The nearest galaxy is far enough ($40'$ or 22 pixels) and does not affect an accuracy of the photometry. 
The robust  ``centroid'' algorithm is used for the background value calculation. The algorithm allows one to exclude the influence of the nearby objects.
The data are corrected for the fluctuations of atmospheric opacity using the ``Astrokit'' program\footnote{developed by V.V. Krushinsky and A.Yu.Burdanov 
(President Yeltsin Ural Federal University).} 
that implements a slightly modified algorithm described in \cite {kit}. This program conducts differential photometry using an ensemble of stars that are close to an object.

\begin{figure}[h!]\centering
\includegraphics[width=0.9\linewidth]{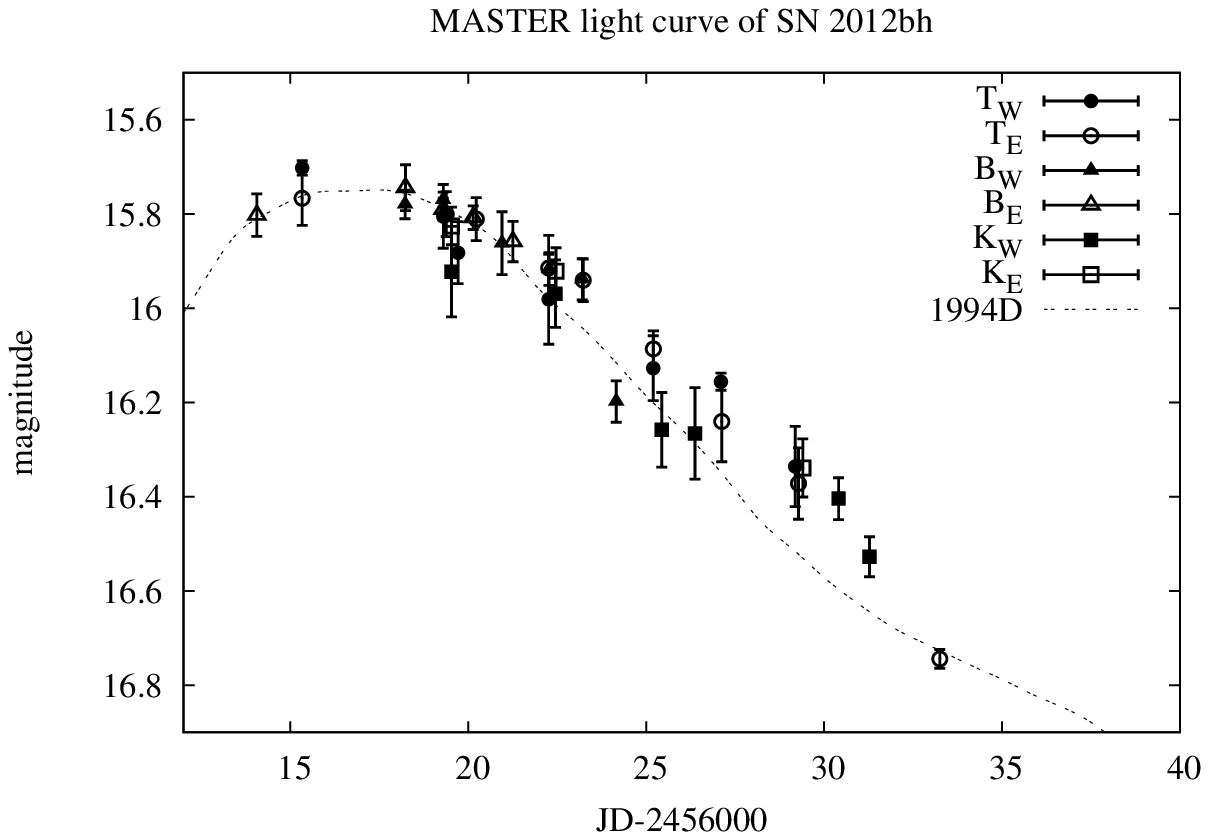}
\caption{Light curve of SN 2012bh recorded by MASTER. Tunka data are shown by circles; Blagoveshchensk, triangles; Kislovodsk, squares. 
Solid and empty signs correspond to different tubes with mutually perpendicular polarizers 
(W -- west, E -- east). The dashed curve is a fit by the light curve of SN 1994D.}
\label{ris:image2} 
\end{figure}

In Table~\ref{tabl:2}, photometry results are presented.
Images with signal/noise more than 30 are summed up for every night.
The combined light curve is presented in Fig.~\ref{ris:image2}. 
For comparison, the light curve of a ``normal'' Type Ia SN 1994D is shown. The light curve of SN 1994D is obtained from its light curves in $B$ and $R$ bands \citep{94D} using relation 
$0.2B+0.8R$. 
Assuming that SN 2012bh is a ``normal'' Type Ia supernova, its light curve can be approximated by the light curve of SN 1994D with parameters:  $T_{max}= 2456017.865$ (March 31), $m_{max}=15.75$.
\begin{table}
\centering
\caption{Photometry results for SN 2012bh by MASTER. $P_{W}, P_{E}$ designate the west and east tubes of telescopes.}\vskip2mm
\begin{tabular}{|c|*{12}{p{0.85cm}|}} 
\hline
\multirow {3}*{JD} & \multicolumn{4}{c|}{Tunka} & \multicolumn{4}{c|}{Blagoveshchensk} & \multicolumn{4}{c|}{Kislovodsk}\\
\cline{2-13}
  & $m_{P_E}$ & $\sigma_{P_E}$ & $m_{P_W}$& $\sigma_{P_W}$& $m_{P_W}$& $\sigma_{P_W}$& $m_{P_E}$& $\sigma_{P_E}$& $m_{P_W}$& $\sigma_{P_W}$& $m_{P_E}$& $\sigma_{P_E}$\\
  &(mag)&(mag)&(mag)&(mag)&(mag)&(mag)&(mag)&(mag)&(mag)&(mag)&(mag)&(mag)\\
\hline
2456014&-&-&-&-&-&-&15.80&0.04&-&-&-&-\\
2456015&15.77&0.06&15.70&0.02&-&-&-&-&-&-&-&-\\
2456018&-&-&-&-&15.78&0.03&15.74&0.04&-&-&-&-\\
2456019&15.80&0.05&15.80&0.07&15.77&0.02&15.79&0.02&15.92&0.10&15.82&0.04\\
2456020&15.81&0.04&-&-&15.86&0.07&15.81&0.02&-&-&-&-\\
2456021&-&-&-&-&-&-&15.86&0.04&-&-&-&-\\
2456022&15.91&0.07&15.98&0.10&15.92&0.03&-&-&15.97&0.07&15.92&0.05\\
2456023&15.94&0.04&-&-&15.94&0.04&-&-&-&-&-&-\\
2456024&-&-&-&-&16.20&0.04&-&-&-&-&-&-\\
2456025&16.09&0.04&16.13&0.07&-&-&-&-&16.26&0.08&-&-\\
2456026&-&-&-&-&-&-&-&-&16.26&0.10&-&-\\
2456027&16.24&0.08&16.16&0.02&-&-&-&-&-&-&-&-\\
2456029&16.37&0.08&16.34&0.08&-&-&-&-&-&-&16.34&0.06\\
2456030&-&-&-&-&-&-&-&-&16.40&0.04&-&-\\
2456031&-&-&-&-&-&-&-&-&16.53&0.04&-&-\\
2456033&16.74&0.02&-&-&-&-&-&-&-&-&-&-\\
\hline
\end{tabular}
\label{tabl:2} 
\end{table}

There is no information about the polarization  of stars in the field around SN 2012bh. 
The standard analysis described in Section 2 is applied.
The assumption of 
small interstellar polarization is quite justified. The supernova is located 70 degrees above the galactic equator and absorption in this direction is very small \citep{ext1,ext2}.  
Based on the data of \cite{mar} we conclude that the ISP doesn't exceed $1\%$. Moreover, according to Serkowski law, the value of galactic interstellar extinction indicates that the ISP in 
this direction is very small $P_{ISP} \le 9E(B-V) = 0.12\%$ \citep{Serkowski1975,Schlegel1998}. 

\begin{figure}[h!]\centering
\includegraphics[width=0.9\linewidth]{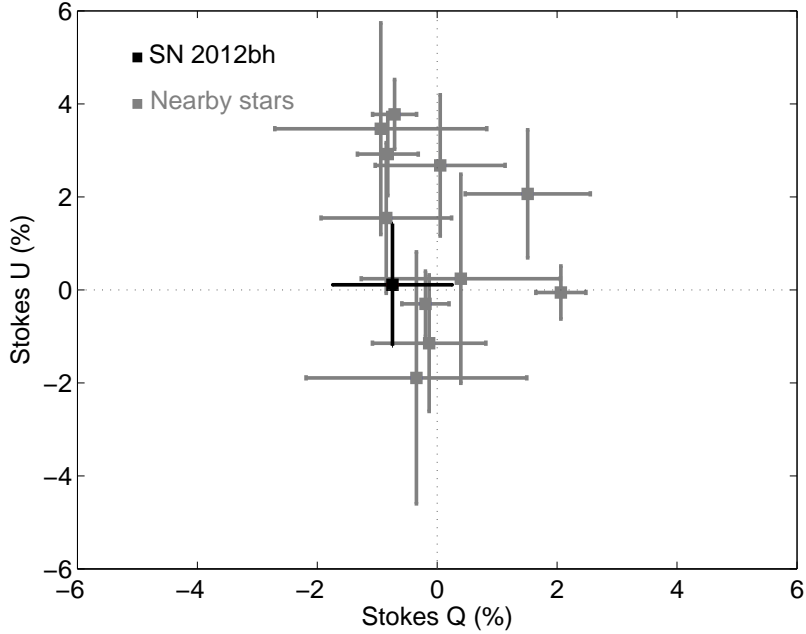}
\caption{QU diagram of the SN 2012bh (black point) and nearby stars (gray points). Stokes parameters are time-averaged results from Blagoveshchensk-Tunka pair of polarizers.}
\label{ris:QU} 
\end{figure}

We derived the polarization degree on different days for two pairs of polarizers: Blagoveshchensk-Tunka and Blagoveshchensk-Kislovodsk. In both cases for every day the polarization 
was near zero. We averaged the polarization degree over all days for each pair of polarizers, then two values obtained were averaged too.
The resulting 1-$\sigma$ upper limit on linear polarization degree of SN 2012bh is 3\%.
QU diagram constructed for Blagoveshchensk-Tunka pair of polarizers is presented in Fig.~\ref{ris:QU}. 

\section{BL Lac objects polarization observations}

MASTER observations of blazars are collected in Table~\ref{table:2}.
For two of them, OC 457 and 3C 454.3, significant polarization was detected. To account for the polarization bias in a case of low signal/noise, we use a traditional statistic correction 
$P_{real}=\sqrt{P^2 - (\sigma_P)^2}$ \citep{Serkowski1958}. The final errors of $P_{real}$ and $\theta$ involve the dispersions of corresponding values of field stars.
Interstellar polarization $P_{ISP}$ is calculated using the empirical law $P_{ISP}\le9E(B-V)$ \citep{Serkowski1975,Schlegel1998}. In all cases it is smaller than 
the dispersion of field stars polarization.
\begin{table}
%\centering
\begin{center}
\caption{List of blazars observed by MASTER in 2012-2013. K~--~Kislovodsk, KB~--~Kislovodsk and Blagoveshchensk.}
\label{table:2} 
\begin{tabular}{|c|c|p{3cm}|c|c|c|}
\hline 
Object  & JD &\centering{Equatorial Coordinates} & $P_{real} (\%)$ & $\theta (^\circ)$ & $P_{ISP} (\%)$\tabularnewline
\hline 
87GB 165943.2+395846 & 2456046.5& $17^h 01^m 24^s.635$ $+39^{\circ} 54' 37''.09$&8$\pm$7&137$\pm$10&0.25\tabularnewline
QSO B1215+303 & 2456047.5& $12^h 17^m 52^s.082$ $+30^{\circ} 07' 00''.64$&4$\pm$2&160$\pm$13&0.2\tabularnewline
OC 457 (K) & 2456331 &$01^h 36^m 58^s.595$ $+47^{\circ} 51' 29''.10$&21$\pm$2&87$\pm$5&1.2\tabularnewline
OC 457 (KB) &2456331 &$01^h 36^m 58^s.595$ $+47^{\circ} 51' 29''.10$&22$\pm$2&92$\pm$4&1.2\tabularnewline
3C 454.3& 2456561.5 &$22^h 53^m 57^s.748$ $+16^{\circ} 08' 53''.56$&34$\pm$2&13$\pm$3&0.8\tabularnewline
\hline 
\end{tabular}
\end{center}
% \label{table:2} 
\end{table}

\begin{figure}[h]
\begin{center}
\begin{minipage}[h]{0.48\linewidth}
\includegraphics[width=0.95\linewidth]{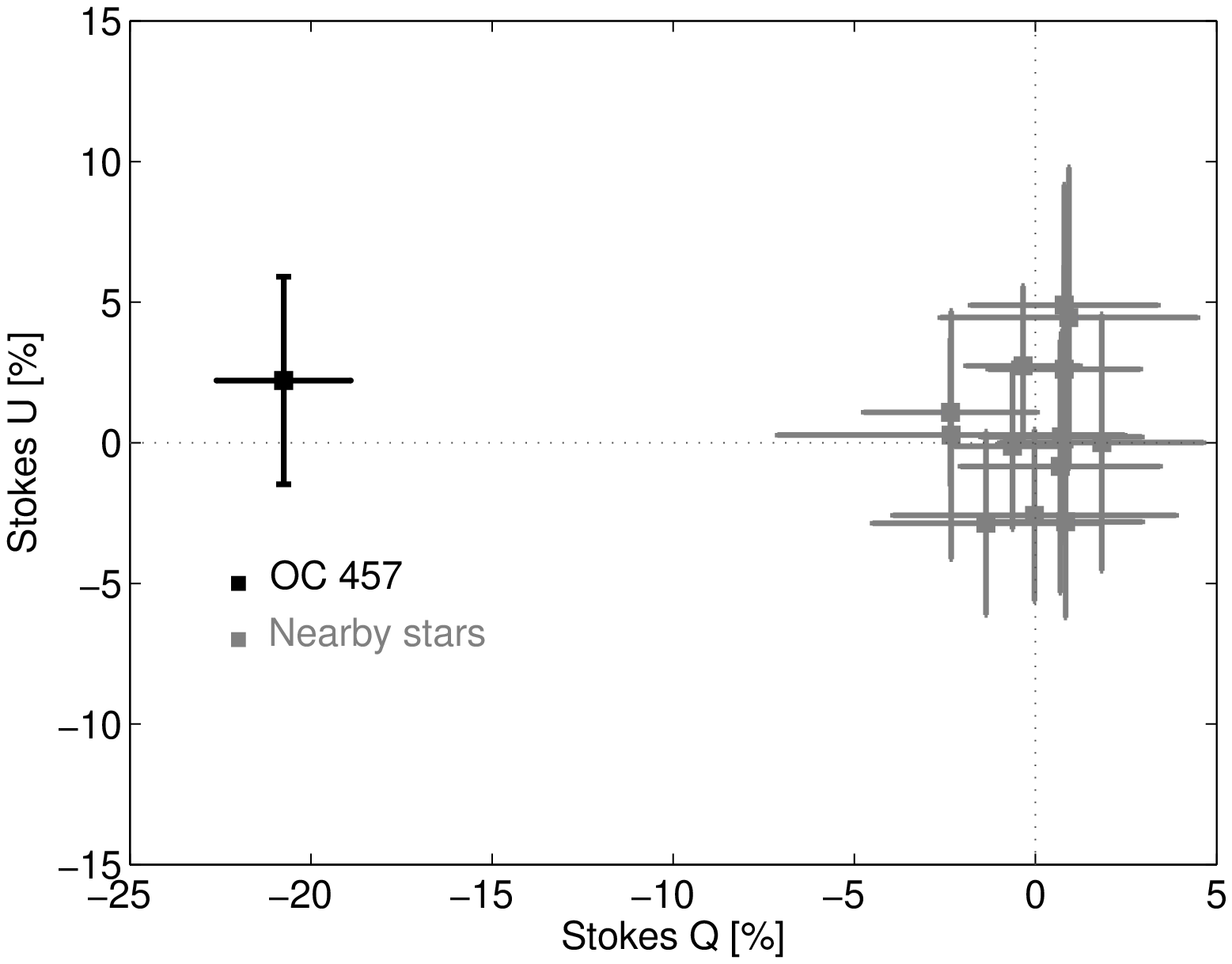}
\caption{QU diagram of the blazar OC 457 (black point) and nearby stars (gray points) observed by MASTER Kislovodsk on February 7, 2013.} 
\label{ris:image5}
\end{minipage}
\hfill
\begin{minipage}[h]{0.48\linewidth}
\includegraphics[width=1\linewidth]{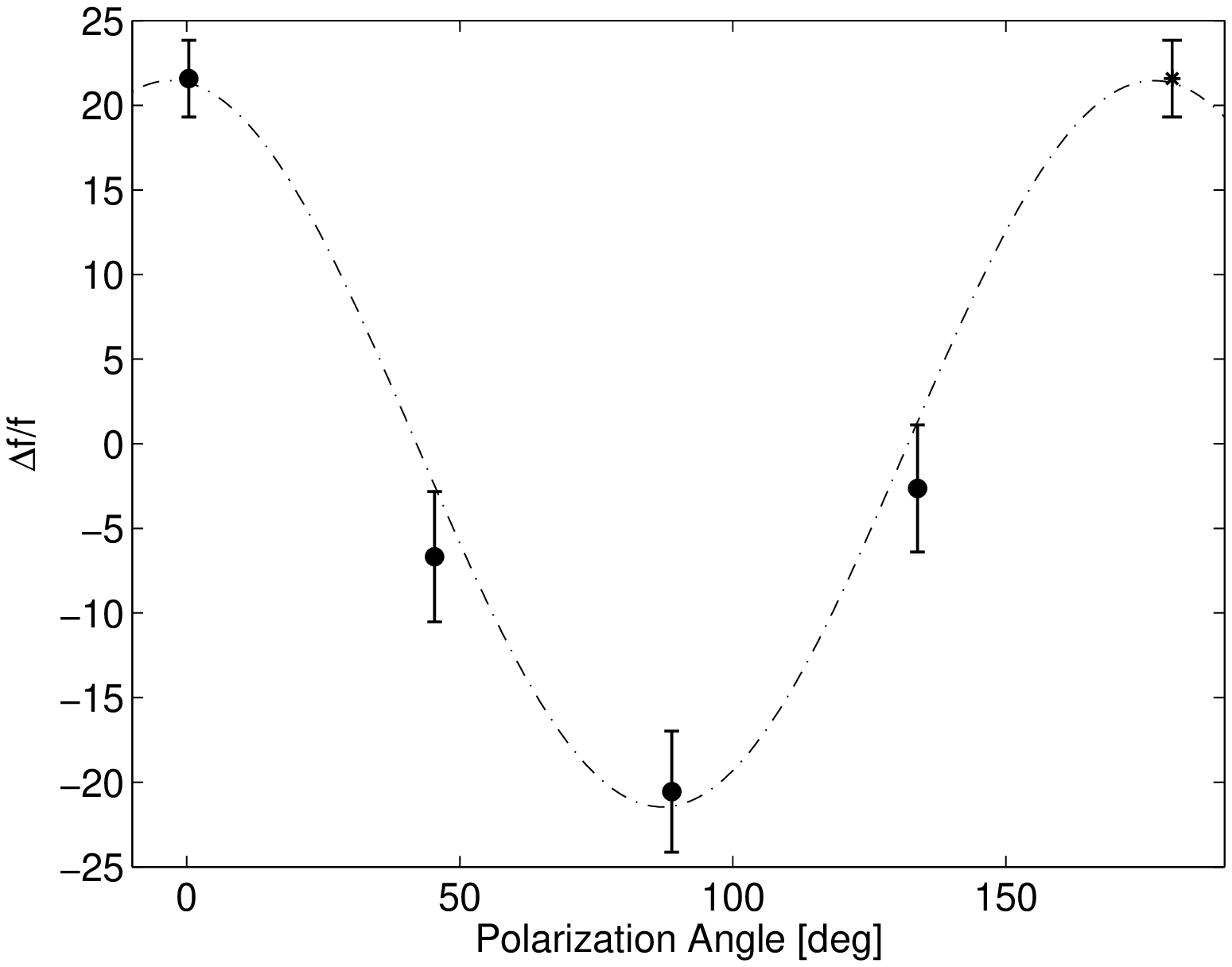}
\caption{Residual flux differences as measured through the polarizers at MASTER Kislovodsk for OC 457. The point at $\theta=0^\circ$ is a repeat of $\theta=180^\circ$. The dashed curve is the $\chi^2$ sinusoid fit.}
\label{ris:image6}
\end{minipage}
\end{center}
\end{figure}

$\bf{OC457}$ In the beginning of 2013 the activity of BL Lac object OC457 increased; in comparison with previous observations its brightness in the R filter went up a fifty-fold \citep{OC457}.
Its redshift is $z=0.859$ \citep{Hall}. Polarization observations were made by MASTER from February 4 to February 7 in Kislovodsk and Blagoveshchensk. Then the average 
magnitude of the object was $15.5^m$ in white light ($0.2B+0.8R$). 
The QU diagram on February 7 is presented in Fig.~\ref{ris:image5}. Using the  measurements with four different position angles, we can find 
the fraction of polarized light with respect to a total intensity for each position angle. These flux differences are shown in Fig.~\ref{ris:image6}. The $\chi^2$ sinusoid fit yields the polarization degree 
and the polarization angle. There is a clear polarization present: $P = (21 \pm 2)\%$, $\theta = (87 \pm 5)^{\circ}$.

\begin{figure}[h]
\begin{center}
\begin{minipage}[h]{0.48\linewidth}
\includegraphics[width=1\linewidth]{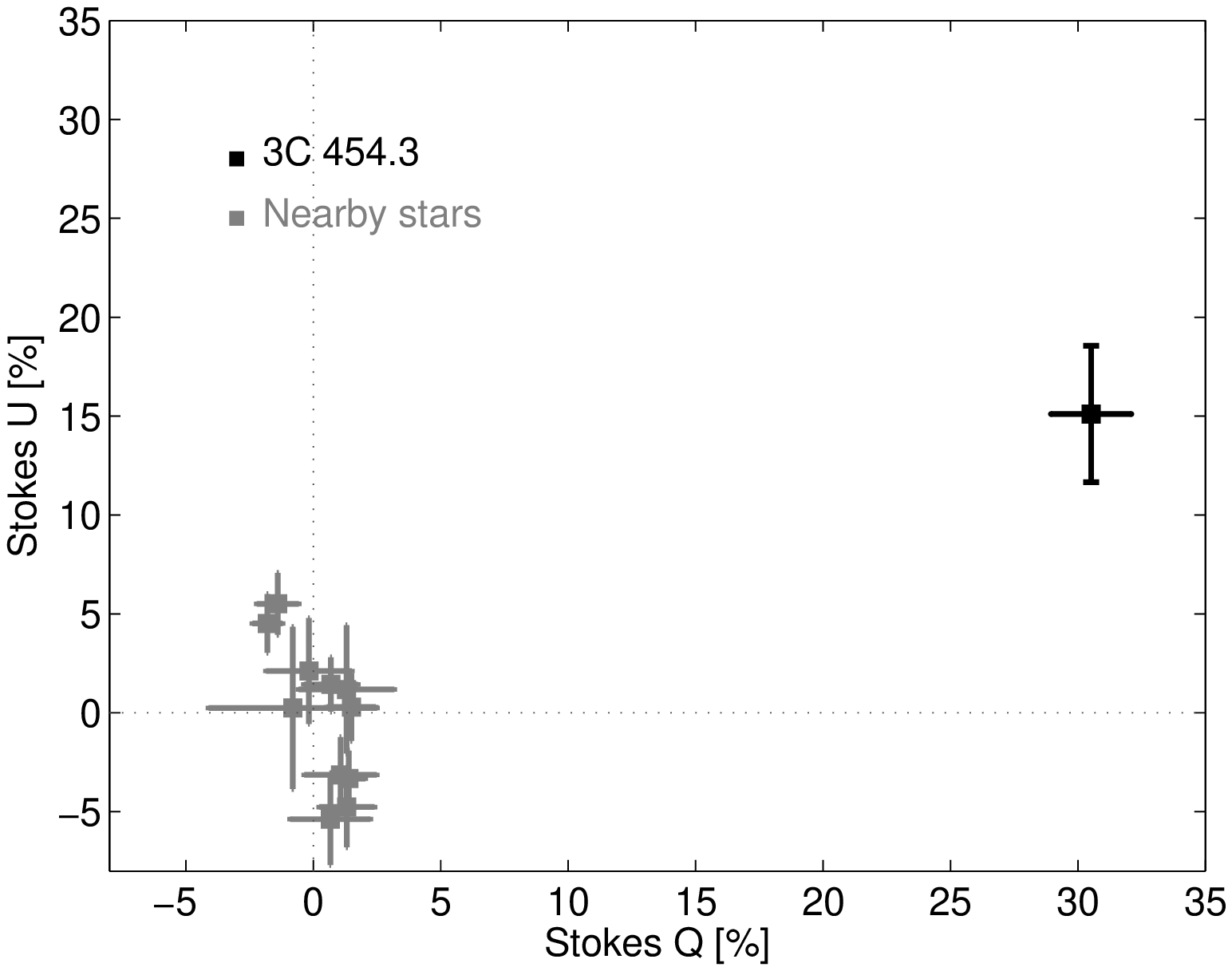}
\caption{QU diagram of the blazar 3C~454.3 (black point) and nearby stars (gray points) observed by MASTER Kislovodsk.}
\label{ris:image7} 
\end{minipage}
\hfill
\begin{minipage}[h]{0.48\linewidth}
\includegraphics[width=1\linewidth]{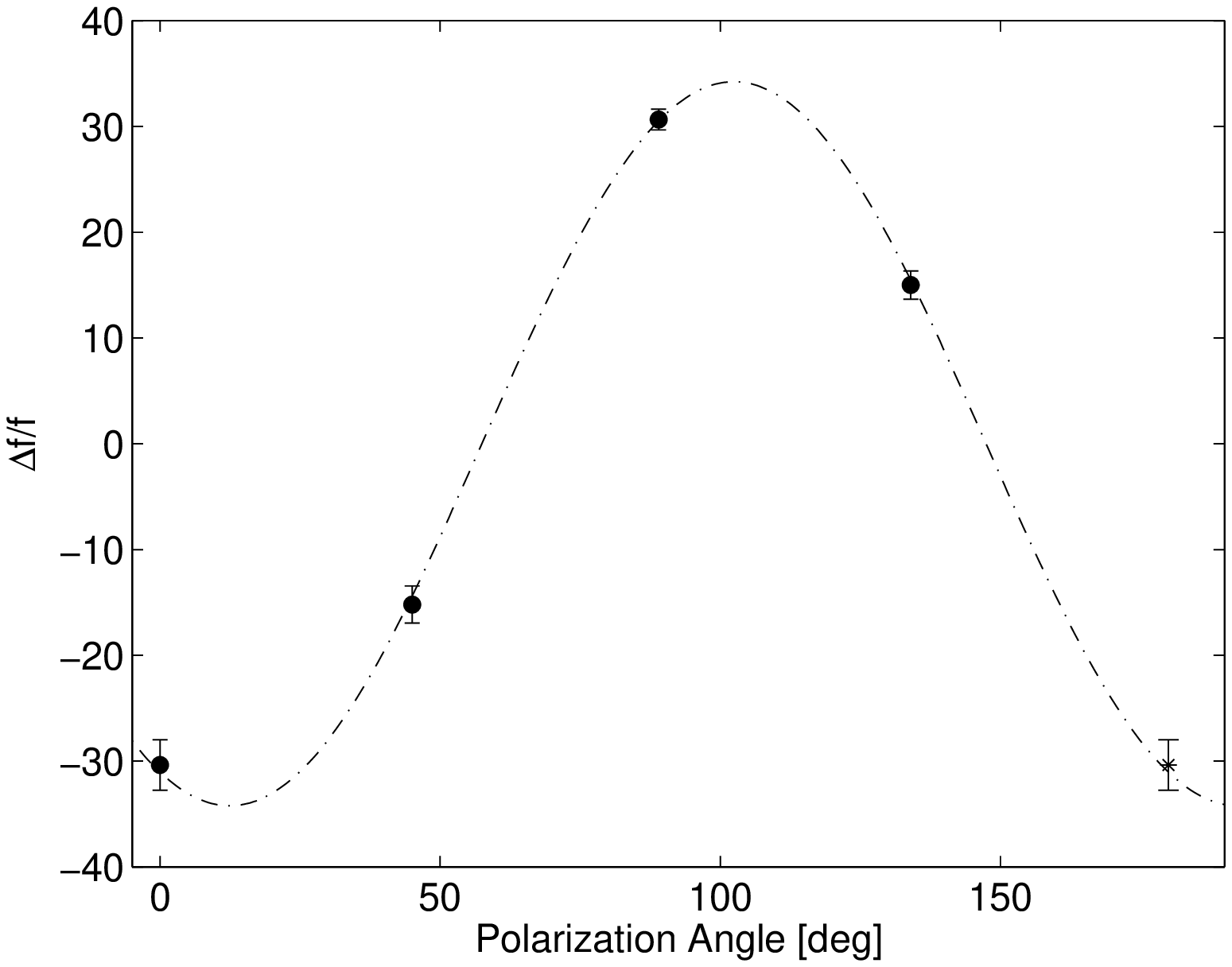}
\caption{Residual flux differences as measured through the polarizers at MASTER Kislovodsk for 3C 454.3. The point at $\theta=0^\circ$ is a repeat of $\theta=180^\circ$. The dashed curve is the $\chi^2$ sinusoid fit.}
\label{ris:image8}
\end{minipage}
\end{center}
\end{figure}

$\bf{3C 454.3}$  On the 24th of September \cite{3C} reported the start of activity of BL Lacertae object 3C 454.3.
MASTER observed it on the night of the 25/26 September, in Kislovodsk, with 4 polarizers.  
The magnitude of the object was $\sim14^m$ in white light. The polarization degree was very high $P = (34 \pm 2)\%$.
The QU diagram and residual flux differences are presented in Figs.~\ref{ris:image7} and ~\ref{ris:image8}.

\begin{figure}[h]
\begin{center}
\begin{minipage}[h]{0.48\linewidth}
\includegraphics[width=1\linewidth]{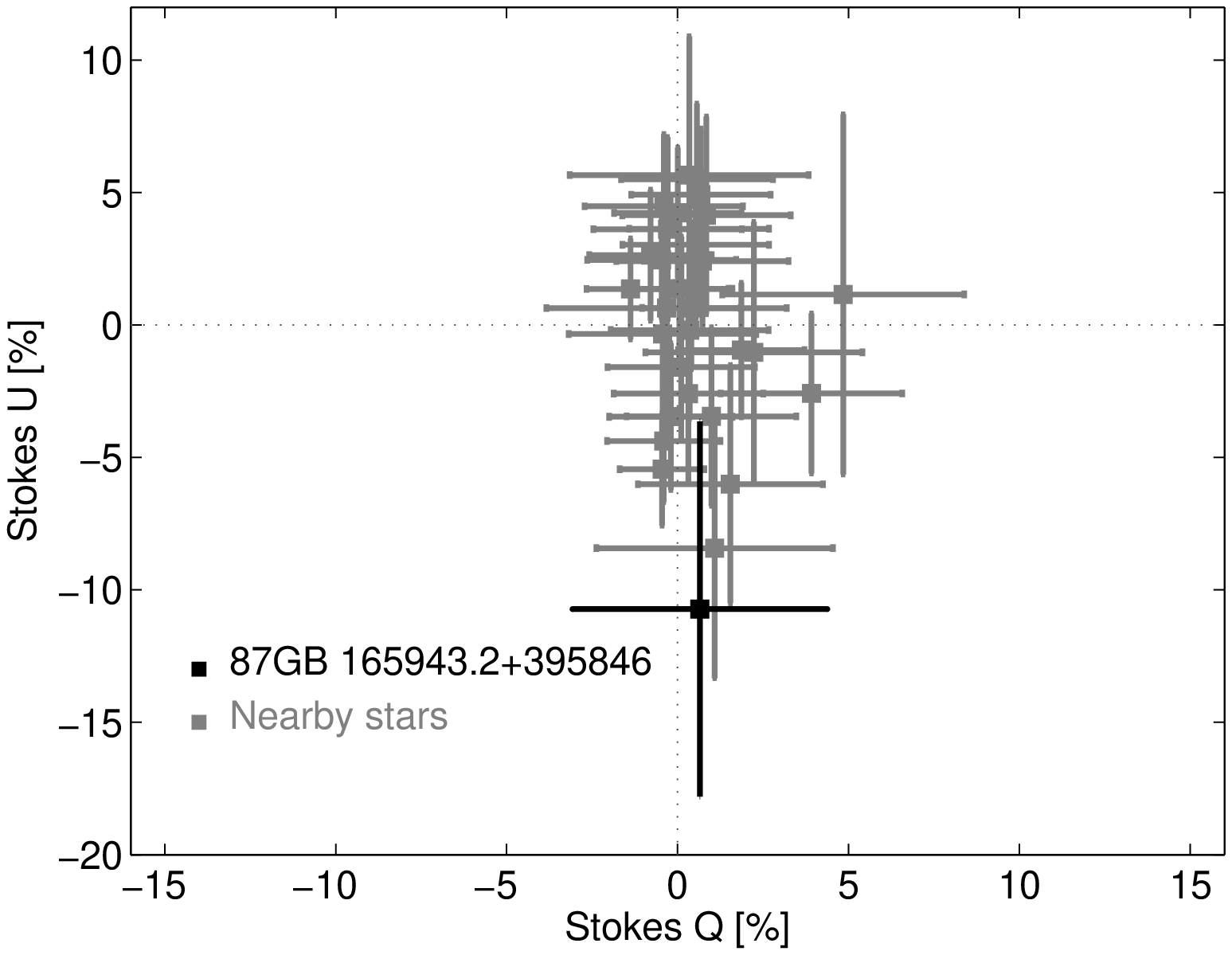}
\caption{QU diagram of the blazar 87GB 165943.2 +395846 (black point) and nearby stars (gray points) observed by MASTER Kislovodsk.} 
\label{ris:image9} 
\end{minipage}
\hfill
\begin{minipage}[h]{0.48\linewidth}
\includegraphics[width=1\linewidth]{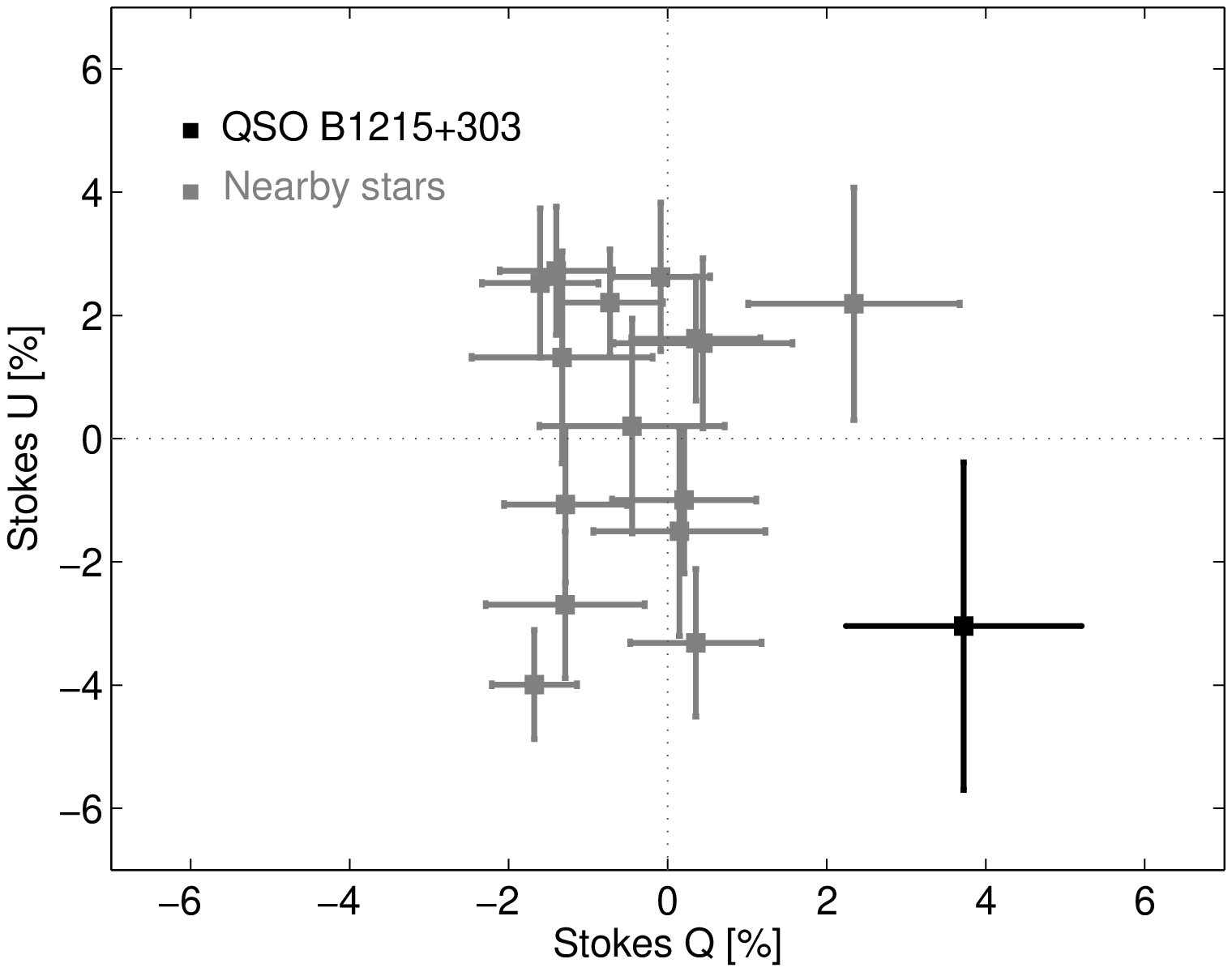}
\caption{QU diagram of the blazar QSO B1215+303 (black point) and nearby stars (gray points) observed by MASTER Kislovodsk.}
\label{ris:image10}
\end{minipage}
\end{center}
\end{figure}

$\bf{87GB 165943.2 +395846}$ $\bf{and}$ $\bf{QSO B1215+303}$ For BL Lac object 87GB 165943.2 +395846 the polarization degree was $P = (8 \pm 7)\%$ at the moment of observations (Fig.~\ref{ris:image9}).
On the night of MASTER observations  BL Lac object 87GB 165943.2 +395846 was fainter than $17^m$ in white light. 
For the BL Lac object QSO B1215+303 the polarization degree was $P = (4 \pm 2)\%$ at the moment of observations (Fig.~\ref{ris:image10}); 
the magnitude of object was $\sim15^m$ in white light. 
Apparently, these blazars were at quiescent states.

The results emphasize once again that MASTER polarizers are able to measure high polarization of bright objects ($<16^m$).

\bigskip

Due to the subtraction of nearby field stars between tubes with differently oriented polarizers MASTER cannot use standard Galactic polarized and unpolarized objects for calibration. 
Blazars are good candidates for calibration of MASTER polarization degree and angle. 
The polarization degree of blazars can reach very high values during their period of activity as well as their brightness. 
On the other hand, blazars are strongly variable sources. For calibration we need to obtain polarimetric data synchronously with other telescopes.

The degree of linear polarization and position angle of blazars OC 457 and 3C 454.3 agree with the results obtained by \cite{Larionov} and the  
data of the virtual observatory of the Laboratory of Observational Astrophysics of the Sobolev
 Astronomical Institute\footnote{http://lacerta.astro.spbu.ru/program.html} at the same observational dates.

\section{Conclusions}
In this paper, we described the goals and methods of polarimetric observations made with the MASTER robotic net. There is a number
of different classes of astrophysical objects that manifest
light polarization. Remarkably, most of them are the sites
of high-energy phenomena. In the paper we present observations of gamma-ray bursts, one supernova, and several blazars.

The polarization observations of blazars show that MASTER's polarizers can be successfully used to measure a linear polarization degree above 5-10\% with position angle accuracy 
of 3--10 degrees depending on brightness. Observing blazars with MASTER benefits also us as a source of calibration for the MASTER polarization measurements.

The polarization measurements are necessary for understanding the
physics of GRBs jets: their geometry, magnetic field, microphysics, and emission mechanisms.
However, GRBs prompt optical emission polarization has not been
registered yet; there are only several measurements of afterglow polarizations.

For GRB100906A, GRB110422A, and GRB121011A, we present the polarization measurements of the prompt optical emission. Unfortunately, in each case, only two orthogonal polarizers took part 
in observations, thus only a lower limit on polarization is deduced and the position angle remains undefined. The dimensionless Stokes parameter is estimated to be less than an observational 
error for single objects, which equals to  2\% in MASTER observations.
We notice that the typical dispersion of  the Stokes parameter of field stars in MASTER observations is 5\%. Thus, MASTER telescopes can safely register linear polarization in excess of 10\% 
and marginally detect polarization just above 5\%. A level of 10\% for the degree  of linear polarization is predicted by some theoretical models of GRB emission.

The supernova phenomena can last several months. It allows us to obtain long data series including polarimetry. The discovery of significant polarizations from Type Ia supernovae will be 
an independent argument for the model of merging white dwarfs as one of the evolutionary path to such supernovae. Moreover, if some of SNe Ia are suspected to be asymmetric, it will 
raise a question about their suitability for cosmology as precise standardizable candles. Recent data shows modest continuum polarization (less than 1\%) of SNe Ia but stronger line 
polarization ($\sim 2\%$).
Low continuum polarization of SNe Ia light could indicate that the explosion is nearly spherical. But theoretical predictions on the polarization
degree from asymmetric explosions are strongly model dependent. Furthermore, observed polarization can be contaminated by interstellar
polarization in our Galaxy and in the host galaxies. It is important to remind that the present number of supernovae with good measurements of polarization
prior to maximum light is still not enough to make a conclusion about the explosion geometry in this class of supernovae.

We present photometry in polarizers of Type Ia SN 2012bh observed by MASTER from March 27 to April 15. The light curve of SN 2012bh looks similar to the light curve of the ``normal'' Type Ia SN 1994D. 
The analysis of the light curve shows that the maximum brightness was on March 31. The resulting 1-$\sigma$ upper limit on the linear polarization degree of SN 2012bh is 3\%.  
 Further spectropolarimetric and polarimetric observations of SNe Ia are needed to investigate explosion geometry and matter distribution 
around supernova and along the line of sight.

\section{Acknowledgements} 
We thank colleagues from the Kourovka Astronomical Observatory, in particular, A.A. Popov, for assistance in photometric processing of some data, and A.Yu. Burdanov, as one of the 
``AstroKit''  program developers. We acknowledge V.M. Larionov and D.A. Blinov for data on blazars and comments. We thank the Extreme Universe Laboratory of Lomonosov Moscow State University Skobeltsyn 
Institute of Nuclear Physics. This work was partially supported by funds from Megagranta No 11.634.31.0076. This work was supported in part 
 by M.V. Lomonosov Moscow State University Program of Development, 
by the Ministry of Education and Science of the Russian Federation (Agreement 8415 on 27 August 2012), State Contract No. 14.518.11.7064 on July 20, 2012 and 
the ``Dynasty'' Foundation of noncommercial programs. G.~V.~Lipunova is supported by RFBR grant 12.02.00186. The work was also supported by RFBR grant 14-02-31546.


\begin{thebibliography}{90}
\bibitem[Abazajian et al.(2009)]{sdss} Abazajian, K.N. et al., 2009, ApJS, 182, 543.
\bibitem[Ahn et al.(2005)]{Seh} Ahn, S.-W. et al., 2005. Nanotechnology 16, 1874.
\bibitem[Barkhouse and Hall(2001)]{Hall} Barkhouse, W. A. \& Hall, P. B., 2001. AJ, 121,  2843.
\bibitem[Barth et al.(2003)]{Barth 2003} Barth, A.J. et al., 2003. ApJ 584, 47.
\bibitem[Bersier et al.(2003)]{Bersier 2003} Bersier, D. et al., 2003. ApJ 583, 63.
\bibitem[Blinnikov et al.(2006)]{94D} Blinnikov, S.I. et al., 2006. A\&A, 453, 229.
\bibitem[Blinov et al.(2013)]{OC457} Blinov, D. et al., 2013. The Astronomer's Telegram No. 4779.
\bibitem[Bonnarel et al.(2000)]{aladin} Bonnarel, F. et al., 2000. A\&AS, 143, 33.
\bibitem[Burstein and Heiles(1982)]{ext2} Burstein, D. \& Heiles, C., 1982. AJ, 87,  1165.
\bibitem[Chornock et al.(2006)]{chornock} Chornock, R., et al., 2006. PASP, 118,  722.
\bibitem[Chornock et al.(2008)]{chornock2006} Chornock, R. \& Filippenko, A.V., 2008. AJ, 136, Issue 6, 2227.
\bibitem[Chornock et al.(2012a)]{PanStar} Chornock, R. et al., 2012. The Astronomer's Telegram No. 3997.
\bibitem[Chornock et al.(2012b)]{CBET3066} Chornock, R. et al., 2012. CBET, No. 3066.
\bibitem[Covino et al.(1999)]{Covino 1999} Covino, S. et al., 1999. A\&A 348, 1.
\bibitem[Covino et al.(2002)]{Covino 2002} Covino, S. et al., 2002. GCN 1498, 1.
\bibitem[Covino et al.(2003)]{Covino 2003} Covino, S. et al., 2003. A\&A 400, 9.
\bibitem[Covino et al.(2004)]{Covino 2004} Covino, S. et al., 2004. ASPC 312, 169.
\bibitem[Evans et al.(2009)]{Evans 2009} Evans, P.A. et al., 2009. MNRAS 397, 1177.
\bibitem[Evans and Marshall(2010)]{Evans 2010} Evans, P.A. \& Marshall, F.E., 2010. GCN 11223, 1.
\bibitem[Everett et al.(2001)]{kit} Everett, M.E. \& Howell, S.B., 2001. PASP, 113, Issue 789,  1428.
\bibitem[Avishay Gal-Yam et al.(2012)]{Atel4293} Gal-Yam, A. et al., 2012, The Astronomer's Telegram No. 4293.
\bibitem[Golenetskii et al.(2011)]{Golenetskii} Golenetskii, S. et al, 2011. GCN 11971, 1.
\bibitem[Gorbovskoy et al.(2012a)]{Gorbovskoy 2012} Gorbovskoy, E.S. et al., 2012. MNRAS 421, 1874.
\bibitem[Gorbovskoy et al.(2012b)]{gcn13854} Gorbovskoy, E.S. et al, 2012. GCN 13854, 1.
\bibitem[Gorbovskoy et al.(2013)]{Gorb} Gorbovskoy, E.S. et al., 2013. Astronomy Reports 57, 233.
\bibitem[G\"otz(2009)]{Gotz 2009} G\"otz, D., 2009. ApJ 695, 208.
\bibitem[Granot(2003)]{Granot} Granot, J. 2003. ApJ 596, 17.
\bibitem[Greiner et al.(2003)]{Greiner 2003} Greiner, J. et al., 2003. Nature 426, 157.
\bibitem[Gress et al.(2011a)]{Gress 2011} Gress, V. et al, 2011. GCN 11960, 1.
\bibitem[Gress et al.(2011b)]{Gress 2} Gress, V. et al, 2011. GCN 12007, 1. 
\bibitem[Gruzinov and Waxman(1999)]{Gruzinov Waxman 1999} Gruzinov, A. \& Waxman, E., 1999. ApJ 511, 852.
\bibitem[Howell et al.(2001)]{howell} Howell, D.A. et al., 2001. ApJ, 556,  302.
\bibitem[Iben and Tutukov(1984)]{DD} Iben, I.Jr. \& Tutukov, A.V., 1984. ApJS, 54, 335.
\bibitem[Kobayashi(2000)]{Kobayashi 2000} Kobayashi, S., 2000. ApJ 545, 807.
\bibitem[Kornilov et al.(2012)]{ExpA} Kornilov, V.G. et al., 2012. Experimental Astronomy 33, 173.
\bibitem[Larionov and Efimova(2013)]{3C} Larionov, V.M. \& Efimova, N.V., 2013. The Astronomer's Telegram No. 5411.
\bibitem[Larionov et al.(2013)]{Larionov} Larionov, V.M. et al., 2013. The Astronomer's Telegram No. 5423.
\bibitem[Lazzati et al.(2003)]{Lazzati 2003} Lazzati, D. et al., 2003. A\&A 410, 823.
\bibitem[Lazzati(2006)]{Lazzati 2006} Lazzati, D., 2006. New Journal of Physics 8, 131.
\bibitem[Leonard et al.(2005)]{leonard} Leonard, D.C. et al., 2005. ApJ, 632,  450.
\bibitem[Lipunov et al.(2004)]{master} Lipunov, V.M. et al., 2004. Astronomische Nachrichten 325, 580. 
\bibitem[Lipunov et al.(2010)]{master1} Lipunov, V.M. et al., 2010. Advances in Astronomy, article id. 349171,~1.
\bibitem[Lipunov et al.(2011)]{lipunov} Lipunov, V.M. et al., 2011. New Astronomy 16, Issue 4,  250.
\bibitem[Lipunov(2012)]{gcn13861} Lipunov, V.M., 2012. GCN 13861, 1.
\bibitem[Malesani et al.(2011)]{Malesani} Malesani, D. et al, 2011. GCN 11977, 1.
\bibitem[Mangano et al.(2011)]{Mangano 2011} Mangano, V. et al, 2011. GCN 11957, 1.
\bibitem[Markkanen(1979)]{mar} Markkanen, T., 1979. A\&A,  74, no. 2,  201.
\bibitem[Markwardt et al.(2010)]{Markwardt 2010} Markwardt, S.D. et al, 2010. GCN 11227, 1.
\bibitem[Masetti et al.(2003)]{Masetti 2003} Masetti, N. et al., 2003. A\&A,  404, 465.
\bibitem[McCall et al.(1984)]{McCall 1984} McCall, M.L. et al., 1984. MNRAS,  210, 839.
\bibitem[Mundell et al.(2007)]{Mundell 2007} Mundell, C.G. et al., 2003. Science 315, 1822.
\bibitem[Ohno  et al.(2012)]{gcn13859} Ohno, M. et al, 2012. GCN 13859, 1.
\bibitem[Palmer et al.(2011)]{Palmer} Palmer, D.M. et al, 2011. GCN 11959, 1.
\bibitem[Patat et al.(2009)]{Patat} Patat, F. et al., 2009. A\&A, 508,  229.
\bibitem[Perlmutter et al.(1999)]{Perlmutter} Perlmutter, S. et al., 1999. AJ 517, 565.
\bibitem[Perna and Lazzati(2002)]{Perna Lazzati 2002} Perna, R. \& Lazzati, D., 2002. ApJ 580, 261.
\bibitem[Postigo et al.(2011)]{Postigo} de Ugarte Postigo, A. et al, 2011. GCN 11978, 1.
\bibitem[Racusin  et al.(2012)]{gcn13845} Racusin, J.L. et al, 2012. GCN 13845, 1.
\bibitem[Riess et al.(1998)]{Riess} Riess, A.G. et al., 1998. AJ 116, 1009.
\bibitem[Rol et al.(2000)]{Rol 2000} Rol, E et al., 2000. ApJ 544, 707.
\bibitem[Rol et al.(2003)]{Rol 2003} Rol, E et al., 2003. A\&A 405, 23.
\bibitem[Sari et al.(1998)]{Sari 1998} Sari, R., Piran, T., Narayan, R., 1998. ApJ 497, 17.
\bibitem[Sari and Piran(1999)]{Sari Piran 1999} Sari, R. \& Piran, T., 1999. ApJ 520, 641.
\bibitem[Schlafly and Finkbeiner(2011)]{ext1} Schlafly, E.F. \& Finkbeiner, D.P., 2011. ApJ, 737,  103.
\bibitem[Schlegel et al.(1998)]{Schlegel1998} Schlegel, D. J., Finkbeiner, D. P. \& Davis, M., 1998, ApJ, 500, 525.
\bibitem[Serkowski(1958)]{Serkowski1958} Serkowski, K., 1958. Acta Astron., 8, 135.
\bibitem[Serkowski(1973)]{Serkowski1973} Serkowski, K., 1973. IAUS no. 52 ''Interstellar Dust and Related Topics'',  145.
\bibitem[Serkowski et al.(1975)]{Serkowski1975} Serkowski, K., Mathewson, D. S. \& Ford, V. L., 1975, ApJ, 196, 261.
\bibitem[Shakhovskoi(1976)]{Shakhovskoi 1976} Shakhovskoi, N.M., 1976. SvAL 2, 107.
\bibitem[Steele et al.(2009)]{Steele 2009} Steele, I.A. et al., 2009. Nature 462, 767.
\bibitem[Tody et al.(1993)]{Tody} Tody, D. et al., 1993. eds, ASP Conf. Ser. Vol. 52, Astronomical Data Analysis Software and Systems II. Astron. Soc. Pac., San Francisco,  173.
\bibitem[Uehara et al.(2012)]{Uehara 2012} Uehara, T. et al., 2012. ApJ 752, 6.
\bibitem[Voshchinnikov(2012)]{ISP} Voshchinnikov, N.V., 2012. Journal of Quantitative Spectroscopy and Radiative Transfer, 113, 2334.
\bibitem[Wang et al.(1996)]{Wang 1996} Wang, L. et al., 1996. ApJ, 467,  435.
\bibitem[Wang et al.(1997)]{wang} Wang, L. et al., 1997. ApJL, 476,  27.
\bibitem[Wang et al.(2003)]{wang2003a} Wang, L. et al., 2003. ApJ, 591,  1110.
\bibitem[Wang et al.(2006)]{wang2006} Wang, L. et al., 2006. ApJ, 653,  490.
\bibitem[Wang et al.(2007)]{Wang 2007} Wang, L., Baade, D. \& Patat, F., 2007. Sci, 315,  212.
\bibitem[Wang and Wheeler(2008)]{Wang and Wheeler} Wang, L. \& Wheeler, J.C., 2008. Annu. Rev. Astron. Astrophys., 46,  27433.
\bibitem[Waxman and Draine(2000)]{Waxman Draine 2000} Waxman, E. \& Draine, B.T., 2000. ApJ 537, 796.
\bibitem[Webbink(1984)]{DD2} Webbink, R., 1984. ApJ 277, 355.
\bibitem[Whelan and Iben(1973)]{SD} Whelan, J. \& Iben, I.Jr., 1973. ApJ 186,  1007.
\bibitem[Xiong  et al.(2012)]{gcn13860} Xiong, S. et al, 2012. GCN 13860, 1.
\bibitem[Yonetoku et al.(2011)]{Yonetoku 2011} Yonetoku, D. et al., 2011. ApJL 743, 30.
\bibitem[Yonetoku et al.(2012)]{Yonetoku 2012} Yonetoku, D. et al., 2012. ApJL 758, 1.
\bibitem[Yurkov et al.(2012)]{gcn13848} Yurkov, V.V. et al, 2012. GCN 13848, 1.
\end{thebibliography}
\end{document}